\documentclass[runningheads,a4paper]{llncs}
\usepackage{subcaption}
\usepackage{xcolor}
\usepackage{hyperref}
\usepackage{tikz}
\usetikzlibrary{arrows,shapes}
\usetikzlibrary{calc}

\usepackage{amsmath}
\usepackage{amssymb}
\usepackage{dsfont}
\usepackage{graphicx}
\usepackage{enumitem}

\usepackage{stackrel}
\usepackage{mathtools}
\usepackage{enumitem}

\usepackage{soul}

\usepackage[dvips]{epsfig}
\usepackage[english]{babel}
\usepackage[T1]{fontenc}
\usepackage[utf8]{inputenc}

\usepackage[algo2e,algoruled,ruled,vlined,linesnumbered]{algorithm2e}

\usepackage{url}

\begin{document}

\newcommand{\refmap}{\curlyvee}
\newcommand{\corefmap}{\curlywedge}

\mainmatter  %
\title{Computing gradient vector fields \\with Morse sequences}

\author{Gilles Bertrand\inst{1}{\tiny\orcidID{0009-0004-7294-7081}}
\and
Laurent Najman\inst{1,2}\tiny{\orcidID{0000-0002-6190-0235}}
}%
\authorrunning{G. Bertrand and L. Najman}%
\institute{Univ Gustave Eiffel, CNRS, LIGM, F-77454 Marne-la-Vallée, France 
\and
Department of Mathematics, Khalifa University, Abu Dhabi, UAE
\email{\{gilles.bertrand,laurent.najman\}@esiee.fr}%
}

\maketitle

\newcommand{\bbbu}{\; \ddot{\cup} \;}
\newcommand{\axr}[1]{\ddot{\textsc{#1}}  \normalsize}

\newcommand{\axcup}{\textsc{C\tiny{UP}} }
\newcommand{\axcap}{\textsc{C\tiny{AP}} }
\newcommand{\axunion}{\textsc{U\tiny{NION}} }
\newcommand{\axinter}{\textsc{I\tiny{NTER}} }

\newcommand{\bb}[1]{\mathbb{#1}}
\newcommand{\ca}[1]{\mathcal{#1}}
\newcommand{\ax}[1]{\textsc{#1} \normalsize}

\newcommand{\axb}[2]{\ddot{\textsc{#1}} \textsc{\tiny{#2}}  \normalsize}
\newcommand{\bbb}[1]{\ddot{\mathbb{#1}}}
\newcommand{\cab}[1]{\ddot{\mathcal{#1}}}

\newcommand{\rel}[1]{\scriptstyle{\mathbf{#1}}}

\newcommand{\rela}[1]{\textsc{\scriptsize{\bf{{#1}}}} \normalsize}

\newcommand{\de}[2]{#1[#2]}
\newcommand{\di}[2]{#1\langle #2 \rangle}

\newcommand{\la}{\langle}
\newcommand{\ra}{\rangle}
\newcommand{\hs}{\hspace*{\fill}}

\newcommand{\cell}{\mathbb{C}}
\newcommand{\cellp}{\mathbb{C}^\times}
\newcommand{\simp}{\mathbb{S}}
\newcommand{\comp}{\mathbb{H}}
\newcommand{\simpp}{\mathbb{S}^\times}
\newcommand{\den}{\mathbb{D}\mathrm{en}}
\newcommand{\ram}{\mathbb{R}\mathrm{am}}
\newcommand{\tree}{\mathbb{T}\mathrm{ree}}
\newcommand{\graph}{\mathbb{G}\mathrm{raph}}
\newcommand{\vertex}{\mathbb{V}\mathrm{ert}}
\newcommand{\edge}{\mathbb{E}\mathrm{dge}}
\newcommand{\equ}{\mathbb{E}\mathrm{qu}}
\newcommand{\esub}{\mathbb{E}\mathrm{sub}}

\newcommand{\topp}{\langle \mathrm{K} \rangle}
\newcommand{\topq}{\langle \mathrm{Q} \rangle}
\newcommand{\topxp}{\langle \mathbb{X}, \mathrm{P} \rangle}
\newcommand{\topxq}{\langle \mathbb{X},\mathrm{Q} \rangle}

\newcommand{\vt}{\mathcal{K}}
\newcommand{\vtp}{\mathcal{T'}}
\newcommand{\vtpp}{\mathcal{T''}}
\newcommand{\vq}{\mathcal{Q}}
\newcommand{\vqp}{\mathcal{Q'}}
\newcommand{\vqpp}{\mathcal{Q''}}
\newcommand{\vk}{\mathcal{K}}
\newcommand{\vkp}{\mathcal{K'}}
\newcommand{\vkpp}{\mathcal{K''}}

\newcommand{\C}{\ensuremath{\searrow^{\!\!\!\!\!C}}}
\newcommand{\Detach}{\ensuremath{\;\oslash\;}}

\newcommand{\mh}[1]{\hat{\textsl{#1}}}

\newcommand{\sig}{\sigma}
\newcommand{\del}{W}

    \newcommand{\ms}{\overrightarrow{W}}
\newcommand{\msi}{\overrightarrow{W_i}}
\newcommand{\msp}{\overrightarrow{W'}}
\newcommand{\mspp}{\overrightarrow{W_p}}

\newcommand{\msim}{\overrightarrow{W_{i-1}}}
\newcommand{\mss}{\widehat{W}}
\newcommand{\msc}{\ddot{W}}
\newcommand{\mso}{\overleftarrow{W}}

\newcommand{\ov}{\overline}
\newcommand{\un}{\underline}
\newcommand{\ha}{\widehat}
\newcommand{\ti}{\widetilde}
\newcommand{\dd}{\ddot}

\newcommand{\ka}{\kappa}
\newcommand{\ova}{\overrightarrow{w}}

\begin{abstract}

We rely on the framework of Morse sequences to enable the direct computation of gradient vector fields on simplicial complexes. A Morse sequence is a filtration from a subcomplex $L$ to a complex $K$ via elementary expansions and fillings, naturally encoding critical and regular simplexes. Maximal increasing and minimal decreasing schemes allow constructing these sequences, and are linked  to algorithms like Random Discrete Morse and Coreduction. 
Extending the approach to cosimplicial complexes
($S=K\setminus L$)
allows for  efficient computation using reductions, perforations, coreductions, and coperforations.
We further generalize to $F$-sequences, which are Morse sequences weighted by an arbitrary stack function $F$, and provide algorithms to compute maximal and minimal sequences. A particular case is when the stack function is given through a vertex map, common in topological data analysis. %
For injective maps, the complex decomposes into lower stars, recovering established methods and enabling parallel computation; for non-injective maps, our approach applies directly without requiring perturbations. Thus, the paper adopts Morse sequences as a framework that simplifies and connects some important existing propagation-based methods, while also introducing new schemes that extend their scope and practical applicability.
\end{abstract}

\section{Introduction}

A fundamental concept in discrete Morse theory~\cite{For98a} is the one of discrete gradient vector field.
We rely on the framework of Morse sequences~\cite{bertrand2025morse} to compute gradient vector fields on simplicial complexes. A Morse sequence is defined as a sequence of simplicial complexes transitioning from a subcomplex $L$ to a complex $K$ through elementary operations: expansions (adding free pairs) and fillings (adding critical simplexes). These sequences naturally yield a gradient vector field, with regular pairs being free and critical simplexes marking topological features.

The aim of this paper is not primarily to introduce algorithms that surpass existing ones in efficiency or in the number of critical simplexes. Rather, our contribution is to adopt the framework of Morse sequences as a pedagogical and coherent way to describe several of the most important propagation-based methods. This perspective provides a simple language in which these techniques can be reformulated and their relationships made clearer. While we do not attempt an exhaustive survey, our presentation helps to organize part of the literature in a structured manner. At the same time, working consistently within this framework also leads to new algorithmic ideas and extensions, which we develop in the sequel of this paper.

In Section~\ref{sec:seq}, we describe two construction schemes: a maximal increasing scheme (building from $L$ to $K$ by prioritizing expansions) and a minimal decreasing scheme (reducing from $K$ to $L$ by prioritizing collapses), linking these to existing algorithms like Random Discrete Morse~\cite{benedetti2014random} and Coreduction~\cite{Mro09}. These approaches align with propagation-based methods that aim to minimize the number of critical simplexes, a key objective in many computational topology applications. In Section~\ref{sec:cosimplicial}, we extend this framework to cosimplicial complexes (sets 
$S=K\setminus L$) and apply operations such as reductions, perforations, coreductions, and coperforations to efficiently compute Morse sequences.

In Section~\ref{sec:Fsequence}, we further generalize Morse sequences to F-sequences, weighted by a stack $F$ (a map on simplexes), ensuring topological consistency across filtration levels. In Section~\ref{sec:maxminseq}, we provide some generic schemes to compute maximal and minimal $F$-sequences, and we give in Section~\ref{sec:algmaxseq} an algorithm for computing a maximal $F$-sequence. (See Appendix~\ref{app:min} for an algorithm that computes a minimal $F$-sequence). 

In Section~\ref{sec:vertexmaps}, we apply our framework to vertex maps, common in topological data analysis. For injective vertex maps, the complex is partitioned into lower stars, enabling parallel computation. In this case, we show that our approach enable us to retrieve established methods~\cite{robins2011theory,fugacci2019computing}.
Thus, our  algorithms offer flexibility for real-world data, by handling non-injective maps directly, without the need of a total order or of a perturbation.

In the conclusion (Section~\ref{sec:conclusion}), 
we highlight that a Morse complex can be constructed while computing a Morse sequence,
and we propose some future research directions.

\section{Morse sequences}
\label{sec:seq}

Let $K$ be a finite family composed
of non-empty finite sets, called {\it simplexes}.
The family $K$ is a {\it (simplicial) complex} if $\sigma \in K$ whenever $\sigma \not= \emptyset$ and $\sigma \subseteq \tau$
for some $\tau \in K$. 
An element of a simplicial complex $K$ is {\it a face of~$K$}.
A {\em facet of~$K$} is a face of $K$ that is maximal for inclusion.
The {\it dimension} of $\sigma \in K$, written $dim(\sigma)$,
is the number of its elements
minus one. If $\sigma \in K$, we write: \\
-  $\partial (\sigma,K) = \{ \nu \in K \; | \; \nu \subseteq \sigma$ and $dim(\nu) =dim(\sigma)-1 \}$, and \\
- $\delta (\sigma,K) = \{ \nu \in K \; | \; \sigma \subseteq \nu$ and $dim(\nu) =dim(\sigma)+1 \}$, \\
which are, respectively, the {\it boundary} and the {\it coboundary of $\sigma$ in~$K$.} \\
A {\it subcomplex of $K$} is a set $L \subseteq K$ which is a simplicial complex.

We recall the definitions of simplicial collapses and simplicial expansions~\cite{Whi39}. \\
Let $K$ be a simplicial complex and let $\sig, \tau \in K$.
The couple $(\sig,\tau)$ is a {\em free pair for~$K$}, 
if $\tau$ is the only face of $K$ that contains $\sig$.
If $(\sig,\tau)$ is a free 
pair for $K$, then the simplicial complex $L = K \setminus \{ \sig,\tau \}$
is {\em an elementary 
collapse of $K$}, and $K$ is {\em an elementary 
expansion of $L$}. 
We say that
$K$ {\em collapses onto $L$},
or that $L$ {\em expands onto $K$},
if there exists a  sequence 
$\langle K=K_0,\ldots,K_k=L \rangle$, such that
$K_i$ is an elementary collapse of $K_{i-1}$, $i \in [1,k]$.

We also recall  the definitions of perforations and fillings \cite{Whi39}. \\
Let $K,L$ be simplicial complexes.
If $\nu \in K$ is a facet of $K$ and if $L = K \setminus \{\nu \}$, we say that
$L$ is {\em an elementary perforation of $K$}, and that
$K$ is {\em an elementary filling of $L$}.

We now introduce the notion of a ``Morse sequence'' by simply considering expansions and fillings
of a simplicial complex \cite{bertrand2025morse}.

\begin{definition} \label{def:seq1}
Let $L\subseteq K$ be two simplicial complexes. A \emph{Morse sequence (from $L$ to $K$)} is a sequence
$\ms = \langle L = K_0,\ldots,K_k =K \rangle$ of simplicial complexes
such that,
for each $i \in [1,k]$, $K_i$ is either an elementary expansion or an elementary filling of $K_{i-1}$.
If $L=\emptyset$, we say that $\ms$ is a \emph{Morse sequence on $K$}.
\end{definition}

Thus, any Morse sequence $\ms$ on $K$ is a \emph{filtration on $K$}, that is a sequence of nested complexes  $\langle \emptyset = K_0,...,K_k =K \rangle$
such that, for each $i \in [0,k-1]$, we have $K_i \subseteq K_{i+1}$; see \cite{Sco19,Edels13}.

Let $\ms = \langle L=K_0,\ldots,K_k=K \rangle$ be a Morse sequence.
We write   
$\diamond \ms$ for the sequence 
$\diamond \ms = \langle \ka_1, \ldots, \ka_k \rangle$ such that, for each $i \in [1,k]$:
\begin{itemize}[noitemsep,topsep=0pt]
\item If $K_i$ is an elementary filling of $K_{i-1}$, then $\ka_i$ is the simplex such that $K_i = K_{i-1} \cup \{\ka_i\}$; we say that
the face $\ka_i$ is \emph{critical for $\ms$}. 
\item If $K_i$ is an elementary expansion of $K_{i-1}$, then
$\ka_i$  is the pair $(\sigma,\tau)$ such that $K_i = K_{i-1} \cup \{\sigma,\tau \}$; we say that
$\ka_i$, $\sig$, $\tau$, are \emph{regular for $\ms$}. 
\end{itemize}
We say that
$\diamond \ms$ is a \emph{simplex-wise (Morse) sequence (from $L$ to $K$)}.
Note that $\diamond \ms$ is a sequence of faces and pairs.

Observe that, if $\ms = \langle \emptyset = K_0,...,K_k = K \rangle$ is a Morse sequence on $K$, with $k \geq 1$, then
$K_1$ is necessarily a filling of $\emptyset$. 
That is, $K_1$ is made of a single face 
(a vertex) that is critical for $\ms$.

\begin{definition} \label{def:seq2}
Let $\ms$ be a Morse sequence. 
The \emph{gradient vector field of 
$\ms$} is the set composed of all regular pairs for $\ms$.
We say that two Morse sequences $\overrightarrow{V}$ and $\ms$
from $L$ to $K$
are \emph{equivalent} if
they have the same gradient vector field.
\end{definition}

Building a gradient vector field from a complex is a fundamental issue in discrete Morse theory. 
It is worth mentioning that using Morse sequences for computing gradient vector fields entails no loss of generality,
see \cite{bertrand2025morse}.

The two following schemes are two basic ways to build a Morse sequence $\ms$ from $L$ to $K$. 
\begin{enumerate} [noitemsep,topsep=0pt]
    \item \emph{The increasing scheme}. We build $\ms$ from the left to the right. Starting from~$L$, we obtain $K$ by iterative expansions and fillings. We say that this scheme is \emph{maximal} if we make a filling only if no expansion can be made. 
    \item \emph{The decreasing scheme}. We build $\ms$ from the right to the left. Starting from $K$, we obtain $L$ by iterative collapses and perforations. We say that this scheme is \emph{minimal} if we make a perforation only if no collapse can be made. 
\end{enumerate}
Clearly, any Morse sequence may be obtained by each of these two schemes whenever the condition of maximality or minimality is not imposed.

The purpose of maximal  and minimal schemes
is to try to minimize the number of critical
simplexes.  This problem is, in general, NP-hard \cite{Jos06}.
Therefore, these methods do not, in general, give optimal results. 
Note that there may exist some differences between the maximal increasing scheme and the minimal
decreasing one. See \cite{fugacci2019computing} and \cite{Bertrand2023MorseSequences} for examples which illustrate this difference. 
 
In the next section we will see that:\\
- There is a link between the  minimal decreasing scheme and the scheme of the algorithm \emph{Random Discrete Morse},
proposed by Benedetti and Lutz in~\cite{benedetti2014random}.  
See also Section 2.3 and  Algorithm 1 in~\cite{Sco19}. \\
- There is a link between the  maximal increasing scheme and the scheme of the algorithm \emph{Coreduction}
proposed by  Mrozek and Batko in \cite{Mro09}. See also \cite{fugacci2019computing} and Algorithm 3.6 in \cite{Har14}.

Note that a distinctive feature of our approach is the use of these schemes to compute Morse sequences.  
A Morse sequence on a complex \( K \) not only defines a gradient vector field on \( K \) but also imposes a specific structure on it.

\section{Cosimplicial complexes}
\label{sec:cosimplicial}

The computation of a Morse sequence from $L$ to $K$, where $L$ and $K$ are simplicial complexes, can be carried out by inductively performing elementary operations 
restricted to the set $S = K \setminus L$.
This set $S$ is not a simplicial complex, but it possesses the following specific structure, that we refer to as a \emph{cosimplicial complex}; see also \cite{Ber25a}
for more details.

\begin{definition} \label{def:cosim0}
Let $S$ be a finite set of simplexes. The set $S$ is a {\em cosimplicial complex} if,
for any $\sig,\tau \in S$, we have $\nu \in S$ whenever $\sig \subseteq \nu \subseteq \tau$. 
\end{definition} 

Note that each simplicial complex is also a cosimplicial complex. \\
Let $S$ be a finite set of simplexes. 
We write $\overline{S}$ for the set of simplexes such that
$\sig \in \overline{S}$ if and only if there exists $\tau \in S$ with $\sig \subseteq \tau$. 
Thus, we have $S \subseteq \overline{S}$. 
We observe that
$\overline{S}$ is a simplicial complex, this complex is the smallest simplicial complex that contains $S$.
It follows that $S$ is a simplicial complex if and only if $\overline{S} = S$.
We write $\underline{S} = \overline{S} \setminus S$.
The following result is straightforward:

\begin{proposition} \label{prop:cosim0}
Let $S$ be a finite set of simplexes. 
The set $S$ is a cosimplicial complex if and only if $\underline{S}$ is a simplicial complex.
\end{proposition}

If $S$ is a cosimplicial complex, then the sets $L = \underline{S}$ and $K = \overline{S}$ are two simplicial complexes such that 
$L \subseteq K$ and $S = K \setminus L$. 

In fact, a set $S$ is a cosimplicial complex if and only if there 
exist two simplicial complexes $L$ and $K$ such that $S = K \setminus L$. This
formulation corresponds to the definition of 
an \emph{open simplicial complex}, 
see \cite{Knu24}.

The following definition is a simple extension of the notion 
of a Morse sequence to cosimplicial complexes.
Proposition \ref{prop:cosim2} provides another way to describe a Morse sequence from $L$ to $K$.
\begin{definition} \label{def:cosim2}
Let $S$ be a cosimplicial complex. 
We say that a sequence $\ms$ is a \emph{Morse sequence
on $S$} if $\ms$ is a Morse sequence from 
$\underline{S}$ to $\overline{S}$. 
\end{definition} 
\begin{proposition} \label{prop:cosim2} Let $(L,K)$ be a pair of simplicial complexes such that $L \subseteq K$,
and let $S = K \setminus L$. A sequence $\ms$ is a Morse sequence from $L$ to $K$ if and only if
$\ms$ is a Morse sequence on $S$.
\end{proposition}

Let $S$ be a cosimplicial complex. 
For each $\nu \in S$,
we write: \\
\hspace*{\fill}
$\partial(\nu,S) = \partial(\nu,\overline{S}) \cap S$ 
and $\delta(\nu,S) = \delta(\nu,\overline{S})$.
\hspace*{\fill} \\
We have $\partial(\nu,S) \subseteq S$ by construction, we also note that $\delta(\nu,S) \subseteq S$.

 We now introduce  four operations which operate only on the set $S$.

\noindent
 Let $S$ be a cosimplicial complex, and 
 let $\sigma,\tau,\nu \in S$, with $\sigma \subseteq \tau$.
 We say that:
 \begin{itemize}[topsep=0cm] 
\item the complex $S \setminus \{ \sigma, \tau \}$ is a \emph{reduction of $S$} if $\delta(\sigma,S) = \{ \tau \}$,
\item the complex $S \setminus \{ \nu \}$ is a \emph{perforation of $S$} if $\delta(\nu,S) = \emptyset$,
\item the complex $S \setminus \{ \sigma, \tau \}$ is a \emph{coreduction of $S$} if $\partial(\tau,S) = \{ \sigma \}$, 
\item the complex $S \setminus \{ \nu \}$ is a \emph{coperforation of $S$} if $\partial(\nu,S) = \emptyset$.
 \end{itemize}
 
\noindent Reductions and perforations are used in the algorithm Random Discrete Morse~\cite{benedetti2014random}; in this algorithm $S$ is a simplicial complex.
Coreductions and coperforations have been introduced in~\cite{Mro09} with the algorithm Coreduction; in this algorithm $S$ is a complex which is more general
than a cosimplicial complex. See also~\cite{fugacci2019computing} and~\cite{Har14}
for other algorithms based on coreductions. The link between these four operations and operations on the simplicial complexes $\underline{S}$ and $\overline{S}$
is the following.

\begin{proposition} \label{prop:cosim3} 
Let $S$ be a cosimplicial complex, and let $\sigma,\tau,\nu \in S$,
with $\sigma \subseteq \tau$. 
\begin{itemize}[topsep=0cm] 
\item $\overline{S} \setminus \{ \sigma, \tau \}$ is a collapse of $\overline{S}$ iff $S \setminus \{ \sigma, \tau \}$ is a reduction of $S$. 
\item $\overline{S} \setminus \{ \nu \}$ is a perforation of $\overline{S}$ iff $S \setminus \{ \nu \}$ is a perforation of $S$.
\item $\underline{S} \cup \{ \sigma, \tau \}$ is an expansion of $\underline{S}$ iff $S \setminus \{ \sigma, \tau \}$ is a coreduction of $S$. 
\item $\underline{S} \cup \{ \nu \}$ is a filling of $\underline{S}$ iff $S \setminus \{ \nu \}$ is a coperforation of $S$. 
 
\end{itemize}
\end{proposition}

Thus, by Propositions \ref{prop:cosim2} and \ref{prop:cosim3}: \\
- A Morse sequence $\ms$ from $L$ to $K$ can be built with the decreasing scheme
by iterative reductions and perforations on the set $S = K \setminus L$. \\
- A Morse sequence $\ms$ from $L$ to $K$ can be built with the increasing scheme
by iterative coreductions and coperforations on the set $S = K \setminus L$.

\section{$F$-sequences}
\label{sec:Fsequence}

In this section, we introduce Morse sequences weighted by a map. %
We first give some basic definitions relative to these maps, see also \cite{Ber25a}.

Let $F$ be a map from a cosimplicial complex $S$ to $\bb{Z}$. We say that $F$ is a {\em stack on $S$} if
we have $F(\sigma) \leq F(\tau)$ whenever $\sigma,\tau \in S$ and $\sigma\subseteq \tau$. 

Let $F$ be a map from a cosimplicial complex $S$ to $\bb{Z}$.
 For any $\lambda \in \bb{Z}$, we write: \\
 \hspace*{\fill}
  $F_\lambda = \{\nu \in S \; \mid \; F(\nu) \leq \lambda \}$ and
$F[\lambda] = \{\nu \in S \; \mid \; F(\nu) = \lambda \}$. \hspace*{\fill} \\
The sets $F_\lambda$ and $F[\lambda]$ are, respectively, {\em the cut and the section of $F$ at level $\lambda$}.

Note that if $K$ is a simplicial complex, %
then the indexed family $(F_\lambda)_{\lambda\in\bb{Z}}$ is a filtration on $K$.

The two following properties are straightforward.
\begin{proposition} \label{prop:stack1}
Let $S$ be a cosimplicial complex and $F$ be a map from $S$ to $\bb{Z}$.
If $F$ is a stack on $S$, then any cut and any section of $F$ is a cosimplicial complex.
\end{proposition}
\begin{proposition} \label{prop:stack2}
Let $K$ be a simplicial complex 
and $F$ be a map from $K$ to $\bb{Z}$.
The map $F$ is a stack on $K$ if and only if 
any cut of $F$ is a simplicial complex.
\end{proposition}

 Now, we extend the notion of a Morse sequence to an arbitrary stack $F$ on a simplicial complex $K$.
 In such a sequence, each expansion is an \emph{$F$-expansion}, that is, an expansion that
preserves the topology of all cuts of $F$. 
See also \cite{bertrand2023discrete}, 
which establishes a connection
between $F$-expansions and watersheds.

\medskip
\hspace*{\fill}
{\it In the sequel of this paper, $L$ and $K$ will denote simplicial complexes.}
 \hspace*{\fill}

 \begin{definition} \label{def:stack1}
Let $F$ be a stack on $K$ and let $L$ be a subcomplex of $K$. Let $\ka = (\sigma,\tau)$ be a free pair for $L$.
We say that $\ka$ is 
\emph{a free pair for $F$} if $F(\sigma) = F(\tau)$. If 
$\ka$ is a free pair for $F$, we say that
$L' = L \setminus \{\sigma,\tau \}$ is an \emph{(elementary) $F$-collapse of $L$}
and $L$ is an \emph{(elementary) $F$-expansion of $L'$}. 
We write $F(\ka) = F(\sigma) = F(\tau)$.
\end{definition}

Let $F$ be a stack on $K$ and let $L$ be a subcomplex of $K$. 
Let $(\sigma,\tau)$ be a free pair for $L$ which is also a  free pair for $F$. 
Then, it can be easily checked that: \\
- For each $\lambda < F(\sigma)$, we have $\sigma \not\in
L \cap F_\lambda$ and $\tau \not\in L \cap F_\lambda$, \\
- For each $\lambda \geq F(\sigma)$, the pair $(\sigma,\tau)$ is a free pair for $L \cap F_\lambda$.

\begin{definition} \label{def:stack2}
Let $\ms$ be a Morse sequence from $L$ to $K$ and let $F$ be a stack on~$K$. 
We say that $\ms$ is a \emph{Morse sequence on $F$ or an $F$-sequence (from $L$ to $K$)}, if we have $F(\sig) = F(\tau)$ whenever 
$(\sig,\tau)$ is a regular pair for $\ms$.
We also say that an $F$-sequence from $\emptyset$ to $K$ is an \emph{$F$-sequence on $K$}.
\end{definition}

Let $\ms = \langle L = K_0,..., K_k = K \rangle$
be an $F$-sequence. Then $\ms$ induces a ``double filtration'': 
the indexed family $(K_i)_{i \in [0,k]}$ is a  filtration where each $K_i$ induces the filtration $(K_i \cap F_\lambda)_{\lambda \in \mathbb{Z}}$.  
Also, $\ms$ induces the sequence $\langle F_0,..., F_k \rangle$ where each $F_i$ is the stack on $K_i$ which is the restriction
of $F$ to $K_i$.  

\section{Maximal and minimal $F$-sequences}
\label{sec:maxminseq}

\begin{algorithm2e*}[tb]
\renewcommand{\algorithmcfname}{Scheme}%

$I := L$; $X := \emptyset$; $\ms := \la L \ra$\; 
\While{$I \not= K$}{
\lIf{there exists $C$, with $I \subseteq C \subseteq K$, such that $C$ is an $F$-expansion of $I$}{$X :=C$} 

\lIf{$X=\emptyset$}{compute a complex $C$, with $I \subseteq C \subseteq K$, such that $C$ is an elementary filling of $I$, and set $X :=C$}  
$\ms := \ms \cdot X$; $I := X$;  
$X := \emptyset$;
}

\caption{ 
computing a sequence $\protect\ms (L,K,F)$ that is maximal for $F$.}
    \label{SchMax}
\end{algorithm2e*}

\begin{algorithm2e*}[tb]
\renewcommand{\algorithmcfname}{Scheme}%

$I := K$; $X := \emptyset$; $\ms := \la K \ra$\;
\While{$I \not= L$}{
\lIf{there exists $C$, with $L \subseteq C \subseteq I$, such that $C$ is an $F$-collapse of $I$}{$X :=C$} 
\lIf{$X=\emptyset$}{compute a complex $C$, with $L \subseteq C \subseteq I$, such that $C$ is an elementary perforation of $I$, and set $X :=C$}  
$\ms := X \cdot \ms$; $I := X$;  
$X := \emptyset$;
}

\caption{ 
computing a sequence $\protect\ms (L,K,F)$ that is minimal for $F$.}

    \label{SchMin}
\end{algorithm2e*}

We extend to $F$-sequences the maximal and minimal schemes presented in Section \ref{sec:seq}.
These schemes may be formalized with the following definition.

Let $F$ be a stack on $K$ and let $\ms = \langle L = K_0,..., K_k = K \rangle$
be an $F$-sequence. \\
For any $i \in [0,k]$, we say that $K_i$ is \emph{maximal} (resp. \emph{minimal) (for $F$)} if
no $F$-expansion (resp. $F$-collapse) of $K_i$ is a subset of $K$ (resp includes $L$). \\
We say that $\ms$ or $\diamond \ms$ is \emph{maximal (for $F$) } if, for any $i \in [1,k]$, the complex $X_{i-1}$ is maximal  for $F$
whenever $X_i$ is critical for $\ms$. \\
We say that $\ms$ or $\diamond \ms$ is \emph{minimal (for $F$)} if, for any $i \in [0,k-1]$, the complex $X_{i+1}$ is minimal  for $F$
whenever $X_i$ is critical for $\ms$. 

Maximal $F$-sequences fit with
the strategy underlying the algorithm {\em Morse-Reduce}~\cite{mischaikow2013morse}, which operates on filtrations of Lefschetz complexes; here, we work with stacks rather than filtrations, and we focus specifically on simplicial complexes. In addition, minimal $F$-sequences can be seen as extending
the construction used in the algorithm
\emph{Random Discrete Morse}~\cite{benedetti2014random} to the setting of stacks. 

\def\figSize{0.33}
\begin{figure*}[tb]
    \centering
    \begin{subfigure}[t]{\figSize\textwidth}
        \centering
        \includegraphics[width=.99\textwidth]{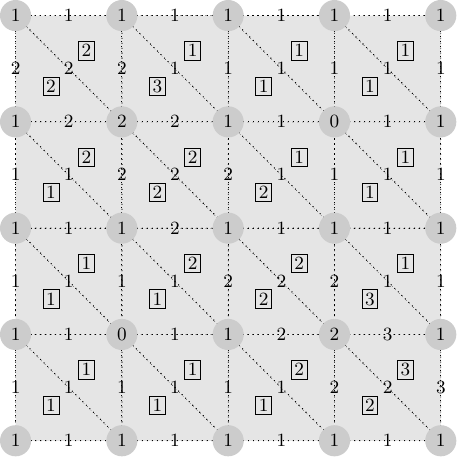}
        \caption{A simplicial stack $F$}
    \end{subfigure}%
\hfill
\begin{subfigure}[t]{\figSize\textwidth}
        \centering
        \includegraphics[width=.99\linewidth]{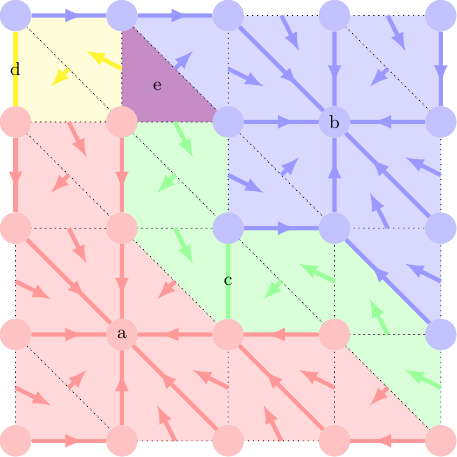}
        \caption{A maximal $F$-sequence}
    \end{subfigure}%
     \hfill 
     \begin{subfigure}[t]{\figSize\textwidth}   \includegraphics[width=.99\textwidth]{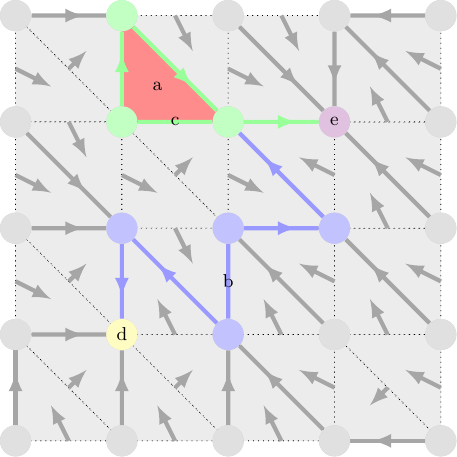}
    \caption{A minimal $F$-sequence}
    \end{subfigure}%
 \caption{(a) A simplicial stack $F$ on a triangulation $K$ of a square. (b) A maximal $F$-sequence on $K$.  (c) A minimal $F$-sequence  on $K$.
 }
 \label{fig:MaxMinFSequence}
\end{figure*}

Some pseudocode for computing a maximal ({\em resp.} minimal) $F$-sequence is provided in Scheme~\ref{SchMax}  ({\em resp.} Scheme~\ref{SchMin}). 

Figure~\ref{fig:MaxMinFSequence}.b illustrates an example of a maximal $F$-sequence
on $K$, where $K$ is a
triangulation of a square, and $F$ is the stack depicted Fig.~\ref{fig:MaxMinFSequence}.a.
Such a sequence begins from either one of the two possible points, both assigned a weight of 0, as shown in Figure~\ref{fig:MaxMinFSequence}.a. Using Scheme~\ref{SchMax} with $L=\emptyset$, and processing simplexes according to their weights, we initiate the sequence with a critical 0-simplex, denoted \textbf{a}. We then perform all feasible expansions with 1-simplices, followed by all possible 2-simplices, resulting in the red region depicted in Figure~\ref{fig:MaxMinFSequence}.b. Next, we introduce a second critical point, \textbf{b}, and repeat the expansion process, yielding the blue region. Subsequently, a critical 1-simplex (an edge), labeled \textbf{c}, is added, and its expansion produces the green region. A further critical 1-simplex, labeled \textbf{d}, is incorporated, leading to the yellow region after expansion. Finally, the sequence is completed by adding a critical 2-simplex, a triangle labeled \textbf{e}.

Fig.~\ref{fig:MaxMinFSequence}.c illustrates an example of a minimal $F$-sequence on the same stack as in Fig.~\ref{fig:MaxMinFSequence}.a. Scheme~\ref{SchMin} begins by performing all possible collapses, shown in gray. Next, it introduces the first critical 2-simplex---the triangle labeled \textbf{a}. This is followed by the introduction of the first critical 1-simplex---the edge labeled \textbf{b} (highlighted in blue). After performing all possible collapses from this step, we obtain the region composed of blue edges. The scheme then introduces a second critical 1-simplex---the edge labeled \textbf{c} (in green)---leading, after further collapses, to the region of green edges. Finally, the scheme terminates with two critical 0-simplices: the points labeled \textbf{e} and \textbf{d}.

\section{An algorithm for computing maximal $F$-sequences}
\label{sec:algmaxseq}
\begin{algorithm2e*}[t]
\KwData{ - A cosimplicial complex $S$ with its operators $\partial$ and $\delta$; and\\
- A stack $F : S \rightarrow \mathbb{Z}$.
The datastructure for $S$ is an array that stores the simplexes according to their increasing dimension and weight: we have, for $1 \leq i < j \leq N = Card(S)$:
$dim(S[i]) \leq dim(S[j])$ whenever $F(S[i]) = F(S[j])$; \\ and $F[S[i]] < F(S[j])$ otherwise.}
\KwResult{$\diamond \ms$, a maximal simplex-wise $F$-sequence on $S$.}

$i:= 1$; $T := \emptyset$; $U := \emptyset$; 
$\diamond \ms := \la \; \ra$;

 \ForAll{$\tau \in S$}{
 
  $\rho(\tau) = Card(\partial(\tau,S))$; 
  
  \lIf{ $\rho(\tau) = 1$}{$U := U \cup \{ \tau \}$}
  }

\While{$i \leq N$}{

\While{$U \not= \emptyset$}{
 Extract $\tau \in U$;

 \If{ $\rho(\tau) = 1$}{
 Find out the simplex $\sigma \in \partial(\tau,S)$ such that $\sigma \not\in T$;
 
  \If{ $F(\tau) = F(\sigma)$}{
 
 $\diamond \ms := \diamond \ms \cdot (\sigma,\tau)$; $T := T \cup \{\sigma,\tau\}$;

 \ForAll{$\mu \in \delta(\sigma,S) \cup \delta(\tau,S)$}{
 
 $\rho(\mu) := \rho(\mu) - 1$; 
 
  \lIf{$\rho(\mu) = 1$}{$U := U \cup \{ \mu \}$}
}
}
}
}
\lWhile{$S[i] \in T$ and $i \leq N$}{ $i:= i+1$}

 \If{$i \leq N$}{
 
 $\sigma := S[i]$; $T := T \cup \{\sigma\}$; $\diamond \ms := \diamond \ms \cdot \sigma$;

 \ForAll{$\tau \in \delta(\sigma,S)$}{$\rho(\tau) := \rho(\tau) - 1$;
 
  \lIf{$\rho(\tau) = 1$}{$U := U \cup \{ \tau \}$}
 
 }
}

}
\Return{$\diamond \ms$}
\caption{{\bf Max}($S,F$)} \label{alg:IC}
\end{algorithm2e*}

We now give a description of an algorithm for computing 
a maximal (simplex-wise) $F$-sequence from $L$ to $K$.
The input of Algorithm~\ref{alg:IC} is the set $S = K \setminus L$ and the map~$F$,
which is restricted to $S$. The output {\bf Max}$(S,F)$ 
is a \emph{maximal $F$-sequence on $S$}, that is, a
maximal $F$-sequence 
from $\underline{S}$ to $\overline{S}$. 
By Proposition \ref{prop:cosim2}, this gives the desired sequence from $L$ to $K$. 
By Proposition \ref{prop:cosim3}, {\bf Max}$(S,F)$ can be computed using operations limited to the set $S$.
Consequently, the sets $\underline{S}$ to $\overline{S}$ are not needed for this computation.

The following facts lead directly  to the soundness of the algorithm;
note that the two sets $T \cup \underline{S}$ and $S \setminus T$ are not explicitly considered in the algorithm. 
\begin{enumerate} [topsep=0cm]
\item At the beginning of the algorithm we have $T \cup \underline{S} = \underline{S}$, and at the end we have $T \cup \underline{S} =  \overline{S}$.
\item    At each step of the algorithm, the set $T \cup \underline{S}$ is a simplicial complex and the set $S \setminus T$ is a cosimplicial complex.
\item If $\nu \in S \setminus T$, we have $\rho (\nu) = Card(\partial(\nu,S) \cap (S \setminus T))$.
\item At line 11, the complex $T \cup \underline{S} \cup \{\sigma, \tau \}$ is an elementary $F$-expansion of $T \cup \underline{S}$. 
\item At line 17, the complex  $T \cup \underline{S} \cup \{\sigma\}$ is an elementary  filling of $T \cup \underline{S}$. 
 \end{enumerate}

Observe that, when $S=K$, 
that is, when $L = \emptyset$, $\underline{S} = \emptyset$, and $\overline{S} = K$, Algorithm~\ref{alg:IC} provides 
a maximal $F$-sequence on the simplicial complex $K$.
This case corresponds to the one given 
Fig.~\ref{fig:MaxMinFSequence}.b.

 \begin{proposition} \label{prop:cosim4}
    If $S$ is a cosimplicial complex, then {\bf Max}$(S,F)$ is a maximal simplex-wise  $F$-sequence on $S$.
\end{proposition}

The complexity of the algorithm depends on the data structure used to access the simplicial complex (see~\cite{fugacci2019computing} for a description of several such data structures.)
As long as we can compute $\partial(.,.)$ and $\delta(.,.)$ in  $\mathcal{O}(d)$ time, where $d$ is the dimension of the complex ({\em e.g.}, for example with a graph, or with cubical complexes and the use of a mask to check the neighborhood of a simplex), the complexity of {\bf Max}$(K,F)$ is $\mathcal{O}(dn)$, where $n$  the number of its simplexes.

In a dual way, we can derive an algorithm for a Morse sequence that is minimal for $F$, see
Algorithm~\ref{alg:DC} in Appendix~\ref{app:min}.

Another important case is when a simplicial complex is decomposed into a disjoint union of cosimplicial complexes. In certain cases, Algorithm~\ref{alg:IC} allows processing each cosimplicial complex independently, possibly in parallel. This fact is used in the next section.

\section{Maximal sequences and vertex maps}
\label{sec:vertexmaps}

In topological data analysis, functions often act on the vertex set of a complex, especially for data from digital images or point clouds~\cite{Edels13}. We now derive an instance of our algorithm for this class of functions.

Let $K$ be a simplicial complex. 
A \emph{vertex map on $K$} is a map  $f : V(K) \rightarrow \mathbb{Z}$,
where $V(K)$ is \emph{the vertex set of $K$}, that is, the set composed of all simplexes $\sigma \in K$ such that 
$dim(\sigma) = 0$. 

Let $f$ be a vertex map on $K$. The map $f$ may be extended to all simplexes of $K$ as follows. 
For any $\tau \in K$, we write: \\
\hspace*{\fill}
 $F(\tau) = \max \{ f(\sigma) \; | \; \sigma \in V(K)$ and $\sigma \subseteq \tau \}$. 
 \hspace*{\fill} \\
Clearly, the map $F$ is a stack on $K$. We say that $F$ is the \emph{stack induced by $f$}. 

We say that a vertex map $f$ on $K$ is a \emph{$\vartheta$-map} if $f$ maps distinct elements to distinct elements, that is, if
$f$ is injective. If $f$ is a $\vartheta$-map, we say that the stack induced by $f$ is a \emph{$\vartheta$-stack}.
Also, if $\sigma_1,...,\sigma_N$ is a total ordering of the vertices of $K$, we say that this ordering is 
the \emph{ordering induced by $f$} if we have $f(\sigma_i) < f(\sigma_j)$ whenever $i < j$. 

\noindent
Let $F$ be a $\vartheta$-stack on $K$. 
If $\sigma \in V(K)$,
the \emph{lower star of $\sigma$} is the set: \\
\hspace*{\fill}
$\hat{\delta}(\sigma) = \{\tau \in K \; | \; \sigma \subseteq \tau$ and $F(\tau) = F(\sigma) \}$. 
\hspace*{\fill} 

\noindent
Since a $\vartheta$-map is injective, we immediately derive the following.

\begin{proposition} \label{prop:LS1}
Let $F$ be a $\vartheta$-stack on $K$. 
For each $\sigma \in V(K)$, we have: \\
\hspace*{\fill}
$\hat{\delta}(\sigma) = \{\tau \in K \; | \; F(\tau) = F(\sigma) \}$. 
\hspace*{\fill} 
\end{proposition}

\noindent
Thus, each lower star is a section of $F$, and each non-empty section of $F$ is a lower star. 
By Proposition \ref{prop:stack1}, it means that each lower star is a cosimplicial complex. 
Also, it means that the set of all lower stars constitutes a partition of the complex $K$.

An $F$-sequence can thus be obtained by treating each lower star independently. Since $F$ is constant on a lower star, we introduce the following.

If $S$ is cosimplicial complex, we write   $\mathds{1}_S$ for the function $S \rightarrow \mathbb{Z}$
such that, for each $x \in S$, we have $\mathds{1}_S(x) = 1$. This $\mathds{1}_S$ is a constant function, 
which is 
trivially a stack on $S$.
If $\ms$ is a Morse sequence on a simplicial complex $K$, we say that $\ms$ or $\diamond \ms$ is {\it maximal} if $\ms$ is maximal for the stack $\mathds{1}_K$. 

Algorithm~\ref{alg:LS} allows to compute an $F$-sequence, where $F$ is a stack on~$K$ induced by a $\vartheta$-map $f$.
The algorithm uses a function ${\bf Max}(S)$ which is a particular case of the function  ${\bf Max}(S,F)$
introduced in the previous section with Algorithm~\ref{alg:IC}. More precisely, we have   ${\bf Max}(S) = {\bf Max}(S,\mathds{1}_S)$.

The output of the algorithm is written ${\bf Max}(V, \hat{\delta})$.
The set $V$ is the vertex set $V(K)$ of the simplicial complex $K$. This set is given by an array 
where the vertices are ordered by the total order induced by $f$. Thus the map $f$ and the stack $F$ are not explicitly used.
The lower star $\hat{\delta}(\sigma)$ of each vertex $\sigma$ is encoded 
in order to fulfill the conditions of Algorithm~\ref{alg:IC}. 

By construction, we have the following:

 \begin{proposition} \label{prop:ls2}
The result ${\bf Max}(V, \hat{\delta})$ of Algorithm~\ref{alg:LS} is a simplex-wise Morse sequence on $K$
that  is maximal for each lower star $\hat{\delta}(\sigma)$, with $\sigma \in V = V(K)$. 
\end{proposition}

In~\cite{robins2011theory}, V. Robins {\em et al.} proposed an algorithm for computing a gradient vector field of a
cubical complex, see algorithm ProcessLowerStars of this paper.
As mentioned in \cite{FlorianiFIM15}, several aspects
``candidate the algorithm to be one of the best, topologically
correct algorithms for computing a Forman gradient''.
It is worthwhile to observe that the core of this algorithm and Algorithm~\ref{alg:LS} are the same.  
See Figures 1 and 2 in~\cite{robins2011theory} which can be used for illustrating Algorithm~\ref{alg:LS},
if we replace simplicial complexes by cubical ones.

In \cite{fugacci2019computing}, U. Fugacci {\em et al.} also proposed  an algorithm for computing a
Forman gradient. The input of the algorithm is a totally ordered set of the vertices
of a simplicial complex which agrees with the input of  Algorithm~\ref{alg:LS}.
Here again,  it may be seen that the core of this algorithm and Algorithm~\ref{alg:LS} are the same. \\
Observe that a total ordering $\sigma_1,...,\sigma_N$ of the vertices of  $K$ induces the $\vartheta$-map $f$ such that $f(\sigma_i) = i$, which itself induces the $\vartheta$-stack $F$ as defined above.

\SetKwFor{ForBy}{for}{do in parallel}{end for}

\begin{algorithm2e*}[tb]
\KwData{A vertex set $V$ of a simplicial complex $K$ with its lower star operator $\hat{\delta}$.
An array $V[i]$, $i \in [1,N]$, stores the vertices according to a total order.
\\}
\KwResult{$\diamond \ms$, a Morse sequence on $K$ which is maximal for each lower star.}

\ForBy{$i \in[1,N]$}{
$S := \hat{\delta}(V[i])$;
$\diamond \msi:= ${\bf Max}$(S)$}

$\diamond \ms := \la \; \ra$\; 
\lFor{$i \in[1,N]$}
{
$\diamond \ms := (\diamond \ms)   \cdot  (\diamond \msi)$%
}

\Return{$\diamond \ms$}

\caption{${\bf Max}(V, \hat{\delta})$}
    \label{alg:LS}
\end{algorithm2e*}

A vertex map $f$ given by real data almost never fulfills the condition of a $\vartheta$-map where all values are distinct.
In~\cite{robins2011theory},  the map $f$ is perturbed with a tie-breaking scheme 
to ensure unique values. An alternative approach to this issue consists simply in computing the Morse sequence ${\bf Max}(K,F)$ with Algorithm~\ref{alg:IC},
where $F$ is the vertex stack induced by $f$.

\section{Conclusion}
\label{sec:conclusion}

In this paper, we have introduced Morse sequences as a simple and pedagogical framework within discrete Morse theory, supporting the construction of gradient vector fields on simplicial and cosimplicial complexes. Defined through elementary expansions and fillings, Morse sequences provide a coherent way to encode the topological structure of a space. Building on this, our maximal and minimal schemes, together with algorithms such as ${\bf Max}(K,F)$ and ${\bf Min}(K,F)$, extend naturally to $F$-sequences and vertex maps, demonstrating their usefulness in applications including digital image analysis and point cloud processing. The alignment with existing methods shows that this perspective is compatible with well-established approaches, while parallel processing of cosimplicial complexes enhances computational efficiency. Beyond these specific contributions, the broader value of our work lies in showing how Morse sequences can serve as a clear and consistent language for presenting several important propagation-based methods, highlighting their connections, and motivating new algorithmic directions. In this sense, Morse sequences provide both a convenient tool for understanding the literature and a promising basis for further developments.

A distinctive strength of Morse sequences lies in their direct connection to \emph{Morse references}~\cite{Bertrand2023MorseFrames,bertrand2025morse}. A Morse reference $\refmap$ is an application that maps each simplex to a set of critical simplices of the same dimension via gradient path parity: a critical simplex $\nu$ belongs to $\refmap(\sigma)$ if and only if the number of gradient paths from $\sigma$ to $±\nu$
is odd. Hence, the Morse reference straightforwardly relates to the  \emph{Morse complex} (also known as the \emph{critical complex}.) The Morse reference $\refmap$ of a Morse sequence $\diamond \ms = \langle \ka_1, \ldots, \ka_k \rangle$ can be computed by traversing the sequence $\diamond \ms$ from left to right. Thus, it is possible, based on an algorithm such as ${\bf Max}(K,F)$, to compute both the gradient vector field and the Morse complex in a single pass. 
Since the Morse complex is itself a central structure in computational topology, it can then be directly leveraged for efficient computation of Betti numbers and persistent homology~\cite{robins2011theory,fugacci2019computing}, which are key tools in topological data analysis and related applications.

An interesting direction for future research is the systematic use of critical kernels~\cite{bertrand2007critical} in the context of parallel algorithms. Critical kernels provide a symmetric and homotopy-preserving way of collapsing complexes, and can therefore serve as natural building blocks for parallel reductions. Integrating them into the Morse sequence framework may yield new n-dimensional strategies   distributing computations across independent substructures, while maintaining global topological consistency (see \cite{guillou2023discrete} for a review of existing strategies). This perspective opens promising avenues for scaling up to very large complexes, such as those arising in 3D imaging or high-dimensional data analysis.

Python code implementing the algorithms of this paper for illustration purpose is available at \url{https://github.com/lnajman/MorseSequences}. All these algorithms can straightforwardly be  extended to cubical complexes of arbitrary dimension.

\bibliographystyle{splncs04}
\bibliography{biblio}

\begin{thebibliography}{10}
\providecommand{\url}[1]{\texttt{#1}}
\providecommand{\urlprefix}{URL }
\providecommand{\doi}[1]{https://doi.org/#1}

\bibitem{benedetti2014random}
Benedetti, B., Lutz, F.H.: Random discrete {M}orse theory and a new library of
  triangulations. Experimental Mathematics  \textbf{23}(1),  66--94 (2014)

\bibitem{bertrand2007critical}
Bertrand, G.: On critical kernels. Comptes rendus de l'Acad{\'e}mie des
  sciences. S{\'e}rie I, Math{\'e}matique  \textbf{1}(345),  363--367 (2007)

\bibitem{Bertrand2023MorseSequences}
Bertrand, G.: Morse sequences. In: International Conference on Discrete
  Geometry and Mathematical Morphology. pp. 377--389. Springer (2024)

\bibitem{bertrand2025morse}
Bertrand, G.: Morse sequences: A simple approach to discrete morse theory.
  Journal of Mathematical Imaging and Vision  \textbf{67}(16),  1--22 (2025)

\bibitem{Ber25a}
Bertrand, G.: Morse sequences on stacks and flooding sequences. In:
  International Workshop on Combinatorial Image Analysis. LNCS, Springer
  (2025), \url{https://arxiv.org/abs/2509.01384}

\bibitem{bertrand2023discrete}
Bertrand, G., Boutry, N., Najman, L.: Discrete {M}orse functions and
  watersheds. Journal of Mathematical Imaging and Vision  \textbf{65}(5),
  787--801 (2023)

\bibitem{Bertrand2023MorseFrames}
Bertrand, G., Najman, L.: Morse frames. In: International Conference on
  Discrete Geometry and Mathematical Morphology. pp. 364--376. Springer (2024)

\bibitem{Edels13}
Edelsbrunner, H., Morozov, D.: Persistent homology: theory and practice. In:
  European Congress of Mathematics. pp. 31--50. European Mathematical Society
  (2013)

\bibitem{FlorianiFIM15}
Floriani, L.D., Fugacci, U., Iuricich, F., Magillo, P.: Morse complexes for
  shape segmentation and homological analysis: discrete models and algorithms.
  Comput. Graph. Forum  \textbf{34}(2),  761--785 (2015)

\bibitem{For98a}
Forman, R.: Morse theory for cell complexes. Adv. Math.  \textbf{134},  90--145
  (1998)

\bibitem{fugacci2019computing}
Fugacci, U., Iuricich, F., De~Floriani, L.: Computing discrete {M}orse
  complexes from simplicial complexes. Graphical models  \textbf{103},  101023
  (2019)

\bibitem{guillou2023discrete}
Guillou, P., Vidal, J., Tierny, J.: Discrete morse sandwich: Fast computation
  of persistence diagrams for scalar data--an algorithm and a benchmark. IEEE
  Transactions on Visualization and Computer Graphics  \textbf{30}(4),
  1897--1915 (2023)

\bibitem{Har14}
Harker, S., Mischaikow, K., Mrozek, M., Nanda, V.: Discrete {M}orse theoretic
  algorithms for computing homology of complexes and maps. Foundations of
  Computational Mathematics  \textbf{14},  151--184 (2014)

\bibitem{Jos06}
Joswig, M., Pfetsch, M.E.: Computing optimal {M}orse matchings. SIAM J.
  Discrete Math.  \textbf{20},  11--25 (2006)

\bibitem{Knu24}
Knudson, K.P., Scoville, N.A.: Discrete {M}orse theory for open complexes
  (2024), \url{https://arxiv.org/abs/2402.12116}, working paper or preprint

\bibitem{mischaikow2013morse}
Mischaikow, K., Nanda, V.: Morse theory for filtrations and efficient
  computation of persistent homology. Discrete \& Computational Geometry
  \textbf{50}(2),  330--353 (2013)

\bibitem{Mro09}
Mrozek, M., Batko, B.: Coreduction homology algorithm. Discrete \&
  Computational Geometry  \textbf{41}(1),  96--118 (2009)

\bibitem{robins2011theory}
Robins, V., Wood, P.J., Sheppard, A.P.: Theory and algorithms for constructing
  discrete {M}orse complexes from grayscale digital images. IEEE Transactions
  on pattern analysis and machine intelligence  \textbf{33}(8),  1646--1658
  (2011)

\bibitem{Sco19}
Scoville, N.A.: Discrete {M}orse Theory, vol.~90. American Mathematical Soc.
  (2019)

\bibitem{Whi39}
Whitehead, J.H.C.: Simplicial spaces, nuclei and m-groups. Proceedings of the
  London mathematical society  \textbf{2}(1),  243--327 (1939)

\end{thebibliography}

\appendix
\section{An algorithm for minimal Morse sequences} \label{app:min}

\begin{algorithm2e*}[tb]
\KwData{- A cosimplicial complex $S$ with its operators $\partial$ and $\delta$; and\\
- A stack $F : S \rightarrow \mathbb{Z}$.
The datastructure for $S$ is an array that stores the simplexes according to their decreasing dimension and weight: we have, for $1 \leq i < j \leq N = Card(S)$:
$dim(S[i]) \geq dim(S[j])$ whenever $F(S[i]) = F(S[j])$; \\ and $F[S[i]] > F(S[j])$ otherwise.}
\KwResult{$\diamond \ms$, a minimal
simplex-wise $F$-sequence on $S$.}

$i:= 1$; $T := \emptyset$; $U := \emptyset$;
$\diamond \ms := \la \; \ra$;

 \ForAll{$\sigma \in S$}{
 $\rho(\sigma) := \text{Card}(\delta(\sigma,S))$;

 \lIf{$\rho(\sig$) = 1}{$U:=U \cup \{\sig\}$}
}

\While{$i \leq N$}{

\While{$U \not= \emptyset$}{
 Extract $\sig \in U$;

 \If{ $\rho(\sig) = 1$}{
 Find out the simplex $\tau$ such that $\tau \in \delta(\sig,S)$ and $\tau \not\in T$;
 
   \If{ $F(\tau) = F(\sigma)$}{
 
 $\diamond \ms := (\sig,\tau) \cdot (\diamond  \ms)$; $T := T \cup \{\sig,\tau\}$;

 \ForAll{$\mu \in \partial(\sig,S) \cup \partial(\tau,S)$}{
 $\rho(\mu) := \rho(\mu) - 1$; 

 \lIf{$\rho(\mu) = 1$}{$U := U \cup \{ \mu \}$}

}

}

}
}

\lWhile{$S[i] \in S$ and $i \leq N$}{ $i:= i+1$}

\If{$i\leq N$}{

 $\tau := S[i]$;  $T := T \cup \{ \tau \}$; $\diamond \ms := \tau \cdot (\diamond \ms)$;  

 \ForAll{$\sig \in \partial(\tau)$}{
 $\rho(\sig) := \rho(\sig) - 1$; 

 \lIf{$\rho(\sig) = 1$}{$U := U \cup \{ \sig \}$}
}}}

\Return $\diamond \ms$\;

\caption{{\bf Min}($S,F$)} \label{alg:DC}

\end{algorithm2e*}

In this appendix, we give an algorithm for computing 
a Morse sequence from $L$ to $K$ that is minimal for $F$. 
The input of Algorithm~\ref{alg:DC} is the set $S = K \setminus L$ and the map $F$,
which is restricted to $S$. 
The output {\bf Min}$(S,F)$ 
is a \emph{minimal $F$-sequence on $S$}, that is, a
minimal $F$-sequence 
from $\underline{S}$ to $\overline{S}$.

The same notations as for Algorithm~\ref{alg:IC} are used. We derive in the same manner the 
soundness of the algorithm. Here again, the sets $\underline{S}$ to $\overline{S}$ are not needed for computing {\bf Min}$(S,F)$.

 \begin{proposition} \label{prop:cosim5}
If $S$ is a cosimplicial complex,
then {\bf Min}$(S,F)$ is a minimal 
simplex-wise $F$-sequence on $S$.
\end{proposition}

\end{document}